\documentclass[prb,showpacs,twocolumn,final,superscriptaddress]{revtex4}

\usepackage{amsfonts,amsmath,amssymb,ifpdf,epsfig,graphicx,color}
\usepackage[bookmarks]{hyperref}

\newcommand{\beq}{\begin{equation}}
\newcommand{\eeq}{\end{equation}}

\begin{document}

\title{The X-ray edge singularity in optical spectra of Quantum Dots}

\author{M. Heyl}
\affiliation{Physics Department, Arnold Sommerfeld Center for Theoretical Physics and Center for NanoScience, \\
Ludwig-Maximilians-Universit\"at, Theresienstrasse 37, 80333 Munich, Germany}
\author{S. Kehrein}
\affiliation{Physics Department, Arnold Sommerfeld Center for Theoretical Physics and Center for NanoScience, \\
Ludwig-Maximilians-Universit\"at, Theresienstrasse 37, 80333 Munich, Germany}
\affiliation{Georg-August-Universit\"at G\"ottingen, Friedrich-Hund-Platz 1, 37077 G\"ottingen}

\begin{abstract}
    In this work we investigate the X-ray edge singularity problem realized in noninteracting quantum dots. We analytically calculate the exponent of the singularity in the absorption spectrum near the threshold and extend known analytical results to the whole parameter regime of local level detunings. Additionally, we highlight the  connections to work distributions and to the Loschmidt echo. 
\end{abstract}

\date{\today}


\maketitle

\section{Introduction}

In condensed matter theory the X-ray-edge singularity constitutes one of the most important paradigms appearing in a variety of different contexts. In the X-ray edge problem one probes the response of a fermionic system, interacting or noninteracting, subject to a sudden local perturbation. Its origin lies in the study of X-ray spectra  of simple metals where it was shown that the absorption or emission of a photon corresponds to the sudden switch on or off of a local potential scatterer embedded in a noninteracting Fermi sea~\cite{Mahan,Nozieres,Ohtaka,Schotte}. Since then X-ray edge physics has been found in a variety of different systems such as Luttinger liquids with impurity~\cite{Gogolin}, Anderson impurity and Kondo models~\cite{Kotani_an,Kita,Affleck,Helmes,Tureci}, resonant tunneling current-voltage characteristics through localized levels~\cite{Matveev,Levitov}, fermionic systems with gapped spectra~\cite{gapped}, decoherence in two level systems~\cite{Segal} or work distributions~\cite{Silva}.

In quantum dot experiments X-ray edge physics has been found in resonant tunneling current-voltage characteristics through localized levels~\cite{Geim} where the I-V curves display edge singularities $I\sim \theta(V-V_0) (V-V_0)^{-\gamma}$~\cite{Matveev,Levitov} as a function of the applied bias voltage $V$ at zero temperature with an exponent $\gamma$ that is determined by the associated local perturbation. At nonzero temperatures $T$ the singularity gets smeared and $I T^{\gamma}$ becomes a universal function of $eV/k_B T$~\cite{Frahm} as has been demonstrated in numerous experiments.~\cite{Frahm,IV_experiments}

In this work we focus on the realization of the X-ray edge problem in noninteracting quantum dots by means of optics experiments. The possibility to tune the system parameters in quantum dots enables to vary the relevant quantity in the X-ray edge problem, the phase shift $\delta$ of the conduction band electrons. We analytically calculate the absorption lineshape near the threshold of a suitably initialized quantum dot at zero temperature extending the known analytical results~\cite{Kotani_an,Kita} to the whole parameter regime of local level detunings. This is an important generalization of x-ray edge physics to an experimentally accessible setup and it constitutes one of the very few examples that allow for exact solutions. We show that the absorption spectrum can be identified with a work distribution~\cite{Talkner} for a local quench in a resonant level model. Moreover, we highlight the connection to the Loschmidt echo that can be related to the Fourier transform of the absorption spectrum~\cite{Peres,Silva}. Thus the presented setup allows for the measurement of the Loschmidt echo in a condensed matter system by means of optical spectra.

The paper is organized as follows. First, we outline the experimental setup that allows to mimic the X-ray edge problem in quantum dots. Then we calculate the absorption spectrum near the threshold by an associated Riemann-Hilbert problem~\cite{dAmbrumenil}. In the end we show the results and point out the relation to work distributions and the Loschmidt echo.

\begin{figure}
\centering
\includegraphics[width=0.8\columnwidth]{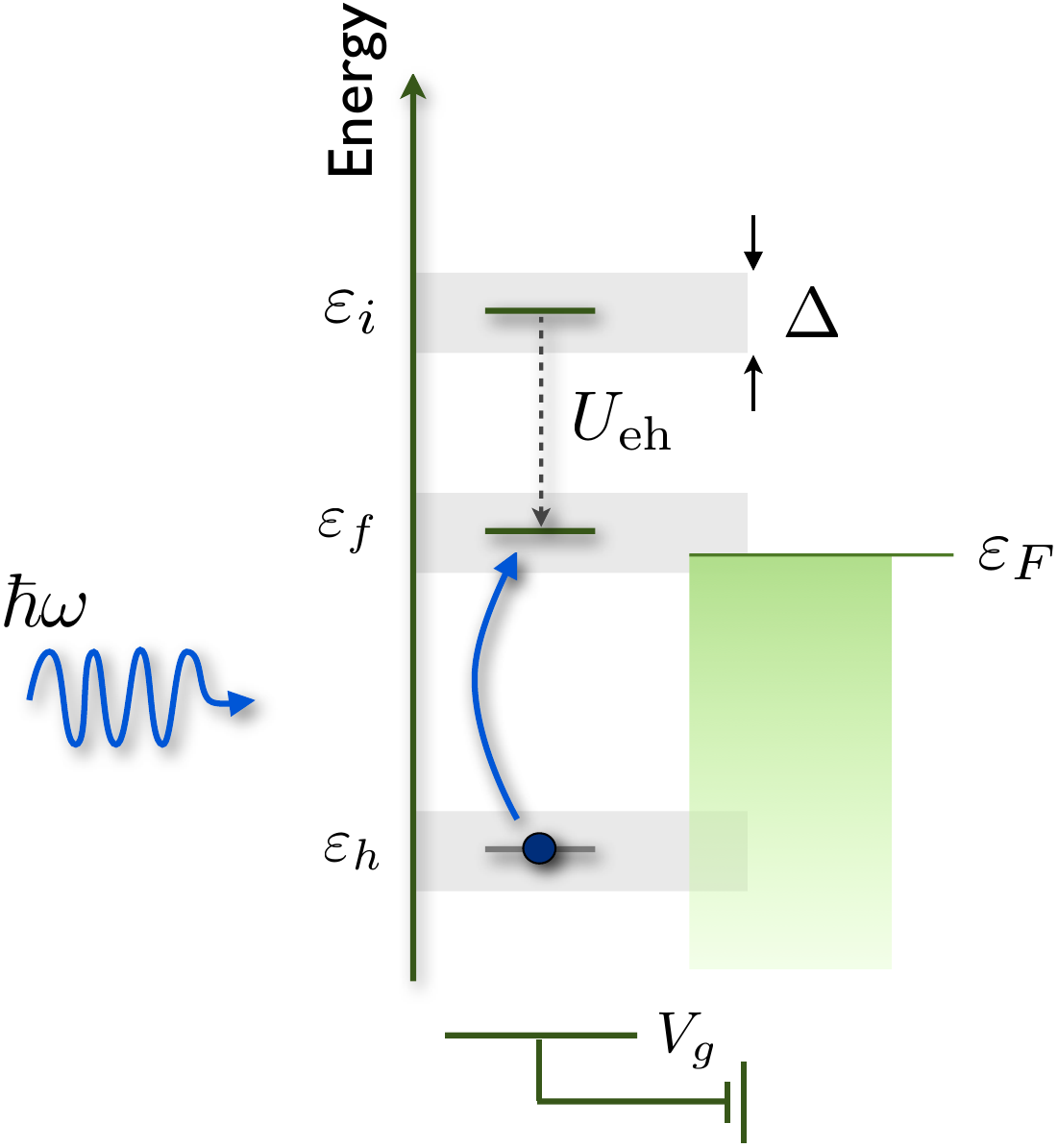}
\caption{(color online) Schematic picture of a quantum dot coupled to a fermionic reservoir that allows to study x-ray edge physics in absorption spectra of quantum dots. The photon absorption of a suitably initialized quantum dot implements a sudden switch on of the tunnel coupling between dot and fermionic reservoir due to a capacitative coupling between the excited electron and the residual hole. For details see text.}
\label{Fig1}
\end{figure}

\section{Modeling the absorption process as a quench in an extended resonant level model}

Below, we present a possible experimental realization of X-ray edge physics in noninteracting quantum dots coupled to an electronic reservoir following the ideas of Helmes \emph{et al.}~\cite{Helmes} and T\"ureci \emph{et al.}~\cite{Tureci}. In Fig.~\ref{Fig1} a schematic picture of the setup is shown. Consider a narrow quantum dot with a large splitting of the single-particle energies. In the following we will assume that the two spin channels are decoupled such that we can restrict to a single channel of spinless fermions. One possible realization of this decoupling is presented in App.~\ref{App_A}. The decoupling of the two spin degrees of freedom eliminates spin fluctuations that can lead to a strongly correlated low-energy state characterized by the Kondo resonance at the Fermi energy in the local density of states. This scenario has been investigated recently in Ref.~\cite{Tureci}. As argued in the App.~\ref{App_A} the formation of a Kondo resonance is avoided in case where charge fluctuations on the quantum dot are sufficiently strong. This can be achieved through a strong coupling between the quantum dot and the conduction band.

By varying the back gate voltage $V_g$ the quantum dot can be tuned in such a way that the topmost occupied level  lies far below the Fermi surface, $(\varepsilon_F-\varepsilon_h)/ \Delta \gg 1$ provided the level splitting is large enough. Here, $\Delta=\pi \rho_0 V^2$ denotes the level broadening with $\rho_0$ the density of states at the Fermi level and $V$ the hopping amplitude of electrons between dot and reservoir. Thus, the lower level can be considered as occupied. If an incident laser beam with angular frequency $\omega$ excites the electron from the lower level into the upper one, a positively charged hole is left behind. Due to a capacitative coupling $U_{\textrm{eh}}$ between the excited electron and the hole the upper level $\varepsilon_i$ is shifted to lower energies $\varepsilon_f$. The localized hole not only interacts with the dot electron, it also establishes a local potential for the conduction band electrons. Assuming that the hole is stable such that it can be considered as static, at least compared to the other time scales in the problem, we can model this system by the following initial (before absorption) and final (after absorption) Hamiltonians:
\begin{eqnarray}
      H_i & = & \sum_k \varepsilon_k \colon c_k^\dag c_k \colon + \varepsilon_i c_d^\dag c_d, \nonumber \\
      H_f & = & \sum_k \varepsilon_k \colon c_k^\dag c_k \colon - g \sum_{kk'} \colon c_k^\dag c_{k'}\colon + \varepsilon_f c_d^\dag c_d + \nonumber \\
	     & &+ V \sum_k \left[ c_k^\dag c_d + c_d^\dag c_k \right] +\Delta E.
\label{eq_Hamiltonians}
\end{eqnarray}
For one particular experimentally relevant realization of these model Hamiltonians, see App.~\ref{App_A}. The hole degree of freedom already has been integrated out and is contained in a constant energy shift $\Delta E$ of the final Hamiltonian. The operator $c_{k}^\dag$ creates an electron with wave vector $k$ in the reservoir. Note that the quantum numbers $k$ refer to an effective one-dimensional chiral description of the electronic degrees of freedom. Thus, we assume $s$-wave scattering which allows for a reduction to a one-dimensional problem. For convenience, the wave vector $k$ is measured relative to $k_F$. The colons $\colon \dots \colon$ denote normal ordering with respect to the Fermi sea. We measure the single particle energies relative to the Fermi level, i.e., $\varepsilon_F=0$. The operator $c_d^\dag$ creates an electron on the upper level of the quantum dot whose energy differs depending on if a photon has been absorbed or not. 

The Hamiltonians in Eq.~(\ref{eq_Hamiltonians}) without the potential scattering term have been introduced in the context of the X-ray edge problem by Kotani and Toyozawa~\cite{Kotani_an,Kotani_num} to describe the X-ray spectra of metals with incomplete shells. They solved the problem analytically in the vicinity of the threshold for the case where the final local level lies far above or below the Fermi energy. Moreover, they phenomenologically inferred from their analytical results the threshold behavior of the absorption spectrum over the whole parameter space. A similar problem at finite temperatures has been investigated in the context of decoherence in charge qubits~\cite{Grishin}. The combined influence of a local potential scatterer and a virtual bound state was first discussed by Kita \emph{et al}.~\cite{Kita} who solved the problem analytically for the case where the final local level energy lies above the Fermi level, i.e., $\varepsilon_f>0$.

The aim of this work is to extend the known analytical zero temperature results to the whole parameter regime of local level detunings with a general framework that can also be useful in other contexts. This includes, for example, decoherence in charge qubits coupled to a defect level~\cite{Grishin,Abel}.

\section{Absorption spectrum}

Assuming that the coupling between the system and the light field is small, one obtains for the absorption spectrum $A(\omega)$, the rate at which photons are absorbed, in second order of the coupling (Fermi's golden rule) at zero temperature
\beq
      A(\omega)=\kappa \sum_n \left| \left\langle e_n \left| c_d^\dag \right| \psi_0 \right\rangle \right|^2 \delta\left[\omega-(e_n-e_{gs})\right].
\eeq
Here, $|\psi_0\rangle$ denotes the ground state of the initial Hamiltonian with energy $e_{gs}$ and $|e_n\rangle$ is a complete orthonormal eigenbasis of the final Hamiltonian with corresponding energies $e_n$. The constant prefactor $\kappa$ contains the experimental details such as the intensity of the incident laser beam and the system-light field coupling. Representing the $\delta$-function by an integral over phase factors one can relate $A(\omega)$ to a dynamical correlation function $G(t)$ via Fourier transformation
\beq
      A(\omega)=\kappa \int \frac{dt}{2\pi} \: e^{i (\omega-\varepsilon_i)t} \: G(t)
\eeq
with
\beq
      G(t)=\langle 0 | e^{iH_i t} e^{-i H_f t} | 0 \rangle.
\label{eq_G}
\eeq
Here, $|0\rangle=c_d^\dag |\psi_0\rangle$ denotes a product state of the Fermi sea for the conduction band electrons with a filled local $d$ orbital. In view of the X-ray edge problem, $G(t)$ is the equivalent to the core-hole Green's function. The dynamical correlation function $G(t)$ in Eq.~(\ref{eq_G}) is an important quantity also in other physical contexts. The quantity $\mathcal{L}(t)=|G(t)|^2$ is the Loschmidt echo that allows to quantify the irreversability of a system~\cite{Peres,Silva}, here $H_i$, under a perturbation, here $H_f-H_i$. Moreover, $G(t)$ is the characteristic funtion of a work distribution $P(\omega)$ for a quench from $H_i$ to $H_f$ where $P(\omega)=\kappa^{-1}A(\omega)$ is the probability of having performed the work $\omega$ on the system under this protocol~\cite{Talkner}. The relation between absorption spectra and work distributions that is evident from a physical point of view has been worked out recently~\cite{Crooks}. A photon when absorbed provides its energy $\omega$ to the system which is equivalent to having performed the work $\omega$.

Analytic results for the dynamical correlation function $G(t)$ in the asymptotic long-time limit $t\to \infty$ have been obtained for the case where the final energy $\varepsilon_f$ of the local $d$ level lies above the Fermi level, i.e., $\varepsilon_f>0$~\cite{Kita}. In the case without potential scatterer, Kotani and Toyozawa~\cite{Kotani_an} calculated analytically the characteristic function $G(t)$ in the limit where the final local energy level lies far above or far below the Fermi level. In both systems, the long-time behavior of the dynamical correlation function $G(t)$ is algebraic
$G(t)\stackrel{t\to \infty}{\longrightarrow} \left( i  \eta t \right)^{-\gamma}, \quad \gamma=\left( 1- \delta/\pi \right)^2$,
with an exponent $\gamma$ that only depends on the phase shift $\delta$ of the conduction band electrons at the Fermi level in presence of the local perturbation. The prefactor $\eta$ of dimension energy is a high-energy scale of the order of the bandwidth. Due to the Friedel sum rule, $\delta/\pi$ is the screening charge that determines the exponent according to the rule of Hopfield~\cite{Hopfield}.

In the following, we will extend the known results to the whole parameter regime including also the case where $\varepsilon_f\leq 0$. Although the problem is in principle quadratic, the mathematical difficulty stems from the fact that in contrast to the original X-ray edge problem an additional \emph{dynamical} degree of freedom, the local $d$ level, and its coupling to the fermionic reservoir is switched on. As a consequence the additional degree of freedom acquires a finite lifetime.

The absorption process creates two local perturbations, the potential scatterer as well as the coupling to a localized level. The time scale for the local level to hybridize with the conduction band is set by the inverse $\Delta^{-1}$ of the equilibrium level broadening $\Delta$. Thus, for times $t\ll \Delta^{-1}$ the local level is effectively decoupled and the dynamics are controlled solely by the potential scatterer. This then leads to the following picture. For times $t \ll W^{-1}$ smaller than the inverse bandwidth $W^{-1}$ the time evolution of $G(t)$ is nonuniversal and is controlled mainly by high-energy excitations. In the intermediate regime $W^{-1} \ll t \ll \Delta^{-1}$ the dynamics is dominated by the local potential scatterer with the local level still effectively decoupled. This is then equivalent to the original X-ray edge problem such that the amplitude $G(t)$ decays algebraically $G(t)\sim(i\eta t)^{-\alpha^2}$ with $\eta$ a high-energy scale of the order of the bandwidth. The exponent $\alpha=\delta^*/\pi$ is set by the phase shift $\delta^*$ for the potential scattering Hamiltonian in Eq.~(\ref{eq_Hamiltonians}) with $V=0$. The dynamics of the system for times $t\gg \Delta^{-1}$ are given by the full Hamiltonian and will be determined via the combined influence of the hybridization as well as the potential scatterer. In the following we will calculate the dynamics in the asymptotic long-time regime $t\gg\Delta^{-1}$ for all local level detunings $\varepsilon_f$ yielding that again $G(t)\sim (i\eta t)^{-(1-\delta/\pi)^2}$ decays algebraically with an exponent that is determined by the phase shift $\delta$.

Due to the quadratic nature of the problem, the final and initial Hamiltonians are both bilinear in fermionic operators, the characteristic function $G(t)$, that is a thermal expectation value of exponentials in $H_i$ and $H_f$, can be reduced to a single-particle problem. Functions such as $G(t)$ can be represented in terms of determinants~\cite{Levitov,Combescot,Klich}
\beq
      G(t)=\mathrm{det}M, \:\: M=1-f+f R,
\eeq
of matrices in the single-particle space due to the Slater determinant structure of the initial state. The matrix $R$ with matrix elements
\beq
      R_{ll'}=\langle |c_l \hat{R} c_{l'}^\dag |\rangle , \:\: \hat{R}=e^{iH_i t} e^{-iH_ft}, \:\: l,l'=k,d,
\eeq
where $|\rangle$ is the true vacuum without any fermion, is essentially determined by the single-particle subspace of $\hat{R}$. The operator $\hat R$ can be idenitified as the time evolution operator of $H=H_f$ in the interaction representation with respect to the free Hamiltonian $H_0=H_i$.  The matrix elements of $R$ reduce to the retarded Green's functions of the final Hamiltonian up to a phase. The initial state is encoded in the matrix $f$:
\beq
      f_{dd}=1,\:f_{dk}=f_{kd}=0,\:\: f_{kk'}=\delta_{kk'}\theta(-k).
\label{eq_Fermi_Dirac}
\eeq
It will be convenient to separate the dynamics of the additional dynamical degree of freedom, the local $d$ level, from the dynamics of the conduction band electrons. For that purpose, we write the matrix $M$ in a block notation
\beq
      M=\begin{pmatrix} A & B \\ C & D \end{pmatrix}
\eeq
where
\beq
      A=M_{dd},\: B_k=M_{dk}, \: C_k=M_{kd},\: D_{kk'}=M_{kk'}
\eeq
such that one obtains by use of an elementary property of the determinant:
\beq
      G(t)=\mathrm{det} M=(A-BD^{-1}C) \: \mathrm{det} D
\label{eq_determinant_formula}
\eeq
where $B D^{-1}C=\sum_{kk'} B_k D^{-1}_{kk'} C_{k'}$ is a scalar. Note that this separation of one degree freedom is formally similar to the treatment of a bound state in the X-ray edge problem in Ref.~\cite{dAmbrumenil}. However, in the present setup the additional $d$ level is a dynamical degree of freedom whereas a bound state is a static object. The matrix $D$ now only includes reservoir states such that $\mathrm{det}D$ can be calculated with techniques known from the original X-ray edge problem. But the separation of the reservoir and $d$ level degrees of freedom comes at the cost of finding the inverse $D^{-1}$ of an infintely large matrix. Using the Riemann-Hilbert method by d'Ambrumenil and Muzykantskii~\cite{dAmbrumenil}, however, the evaluation of the determinant of $D$ is equivalent to finding its inverse $D^{-1}$. In the context of the response of a fermionic system subject to a local perturbation, the auxiliary Riemann-Hilbert problem first appeared in Ref.~\cite{Adamov_Muzykantskii}. Later, it has been used in the theory of full counting statistics~\cite{Muzykantskii_Adamov} and for the X-ray edge problem~\cite{dAmbrumenil,Braunecker} even under nonequilibrium conditions~\cite{FES_nonequilibrium}. In the context of quantum inverse scattering problems the Riemann-Hilbert problem is a well-established technique for evaluating determinants~\cite{quantum_inverse}.

The inversion of the matrix $D$ cannot be done exactly, but only asymptotically for large times $t\gg \Delta^{-1}$. For details, see Ref.~\cite{dAmbrumenil}. In this asymptotic limit it is well known that only the low-energy excitations in the vicinity of the Fermi level are relevant for the dynamics. Assuming that the scattering matrix $S(E)$ for the conduction band electrons in presence of the local perturbation is only weakly dependent on energy one can approximate $S(E)$ by its value at the Fermi level $S(E)\approx S(E_F)=e^{2i\delta}$. Here, $\delta$ is the corresponding phase shift. Within this approximation, the inversion of the matrix $D$ is then equivalent to solving a singular integral equation with a Cauchy kernel~\cite{dAmbrumenil}. Such singular integral equations can be solved analytically due to their relation to Riemann-Hilbert problems~\cite{dAmbrumenil,Mushkelishvili}. For the long-time limit of the generating funtion $G(t)$ one thus obtains
\beq
      G(t)\stackrel{t \gg \Delta^{-1}}{\longrightarrow} (i\eta t)^{-\gamma}, \: \gamma=(1-\delta/\pi)^2.
\label{eq_G_new}
\eeq 
in agreement with the known results for the case $\varepsilon_f>\varepsilon_F$ and consistent with the Hopfield rule of thumb~\cite{Hopfield}. Thus, the known asymptotic behavior extends to the whole parameter regime as already shown in numerous numerical calculations~\cite{Kita,Kotani_num,Oliveira}. This result constitutes one of the rare cases where it is possible to obtain exact analytical solutions.

Eq.~(\ref{eq_G_new}) expresses the asymptotic behavior of the generating function $G(t)$ in terms of the parameters $\eta$ and $\delta$. The quantity $\eta $, a high-energy scale of the order of the bandwidth, cannot be obtained analytically as is usually the case for all the analytical treatments of the X-ray edge problem~\cite{Mahan,Nozieres,Ohtaka,Schotte}. The phase shift $\delta$ of the electrons at the Fermi level is a nonuniversal quantity that depends on a lot of details such as the full free fermionic dispersion relation. Thus, in general it can only be determined numerically for a given system. Only in special cases it is possible to arrive at general statements about $\delta$, for the case of a weak potential scatterer, for example, see Ref.~\cite{Schotte}, for a far detuned local level $|\varepsilon_f-\varepsilon_F|\gg\Delta$ see Ref.~\cite{Kotani_an}. If the final local level energy is resonant with the Fermi level, i.e., $\varepsilon_f=\varepsilon_F$, we have $\delta/\pi=1/2$.

In the context of the initial problem, a quantum dot subject to a laser field, we have to bear in mind that the system actually exhibits two spin channels. If the laser excites electrons of both spins, the problem is still separabel in the spin degree of freedom, i.e., $G(t)=G_\uparrow(t) G_\downarrow(t)$, and the dynamics of each spin component is goverend by the Hamiltonians in Eq.~(\ref{eq_Hamiltonians}). The exponent $\gamma$ of the asymptotic long-time decay of the generating function $G(t)$ gets contributions from both spin channels, i.e., $\gamma=\gamma_\uparrow+\gamma_\downarrow$ with $\gamma_\sigma=(1-\delta_\sigma/\pi)^2$ and $\delta_\sigma$ is the phase shift of the spin-$\sigma$ electrons at the Fermi level. If the incident laser beam is circularly polarized it is possible to address just one of the two electronic spin species, spin-$\uparrow$ say. In this case, only one spin-$\uparrow$ electron is excited from the core-hole into the upper local level. Again, the total exponent $\gamma=\gamma_\uparrow+\gamma_\downarrow$ is given by two contributions. For $\gamma_\uparrow=(1-\delta_\uparrow/\pi)^2$ we then get the same result as in Eq.~(\ref{eq_G_new}). The spin-$\downarrow$ contribution, however, is different as the absorption process does not excite a spin-$\downarrow$ electron in the dot. Thus, we get an exponent $\gamma_\downarrow=(\delta_\downarrow/\pi)^2$ due to the presence of the local potential scatterer generated by the absorption of the spin-$\uparrow$ electron.

\begin{figure}
\centering
\includegraphics[width=\columnwidth]{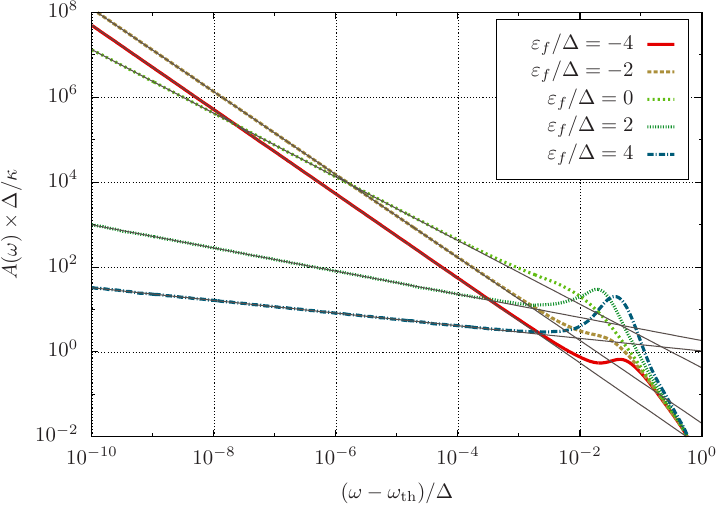}
\caption{(color online) Absorption spectrum $A(\omega)$ as a function of the incident light frequency $\omega$ near the threshold frequency $\omega_{\mathrm{th}}$ for different final energies $\varepsilon_f$ of the quantum dot level at zero temperature. For simplicity we restrict to the case $g=0$ without potential scatterer. Here, $\Delta=\pi \rho_0 V^2$ denotes the half width of the hybridized level in the quantum dot with $\rho_0$ the noninteracting density of states at the Fermi level. The thick lines have been obtained by NRG calculations~\cite{Weichselbaum}. The thin lines show the analytic power-law results that fit perfectly the exact NRG data in the asymptotic low frequency regime for $|\omega-\omega_{\mathrm{th}}|\ll \Delta$. For details, see main text.}
\label{Fig2}
\end{figure}

\emph{Absorption lineshape.} From Eq.~(\ref{eq_G_new}), one can deduce the behavior of the absorption lineshape near the threshold analytically
\beq
      A(\omega)\stackrel{\omega \to \omega_{\mathrm{th}}}{\sim} \theta(\omega-\omega_{\mathrm{th}}) \left( \omega-\omega_{\mathrm{th}} \right)^{\gamma-1}
\eeq
that shows the typical power-law singularity. The singularity is a consequence of the singular behavior of the initial Fermi-Dirac distribution of the conduction band electrons at zero temperature. Thus, at non-zero temperatures $T$ the singularity is cut off~\cite{Anderson_Yuval}, see Ref.~\cite{Braunecker} for the finite temperature generalization in the context of the Riemann-Hilbert method. In Fig.~\ref{Fig2}, NRG data for the absorption spectrum is shown. For light frequencies $\omega$ in the vicinity of the threshold, the analytical power-law results included as thin solid lines fit perfectly to the exact NRG results. The analytical curves in Fig.~\ref{Fig2} are obtained by a fit of the high-energy scale $\eta$ that cannot be obtained analytically by the present appoach as mentioned before. The phase shift $\delta$, however, is not fitted, it is rather obtained within NRG independently of the absorption spectrum.

\emph{Work distribution.} In view of the equivalence to a work distribution, the existence of the threshold in the absorption spectrum is evident. In the beginning, the system is prepared in the ground state of the initial Hamiltonian. The minimum energy, i.e., work, that has to be provided to the system by switching on the coupling to the resonant level is the ground state energy difference between initial and final Hamiltonian. Thus, it is impossible for a photon of energy less than the ground state energy difference to be absorbed. The singular behavior of the absorption spectrum shows that the dominant excitations that are created by the absorption process are low-energy excitations in the vicinity of the Fermi level.

\emph{Loschmidt echo.} As already mentioned before, the characteristic function $G(t)$ is also related to the Loschmidt echo~\cite{Peres,Silva}
\beq
      \mathcal{L}(t)=|G(t)|^2=\left| \left\langle 0\left| e^{iH_i t} e^{-i H_f t} \right| 0 \right\rangle \right|^2.
\eeq
The Loschmidt echo quantifies the stability of motion in time of a system, in this case the Hamiltonian $H_i$, under a perturbation $H_f-H_i$. Thus, for long times $t$ Eq.~(\ref{eq_G_new}) states that, no matter how small the local perturbation is, the time evolution of the state $|0\rangle$ with the final Hamiltonian drives the system into a subspace of the Hilbert space that is orthogonal to the initial state. From the Anderson orthogonality catastrophe~\cite{Orthogonality} it is known that the ground state of the final Hamiltonian is contained in this subspace. The system as a whole, however, does not evolve into the ground state of the final Hamiltonian $|0_f\rangle$ as the overlap of both wave functions $|\langle 0_f |e^{-i H_f t} |0\rangle|^2=|\langle 0_f |0\rangle|^2\sim N^{-\alpha^2}$ with $\alpha=\delta/\pi$  is constant in time. Here, $N$ is the particle number. Thus, the vanishing behavior of $G(t)$ for $t\to \infty$ cannot be simply traced back to the Anderson orthogonality catastrophe~\cite{Orthogonality}, i.e., the vanishing overlap between ground state wave functions. For the original X-ray edge model including only the potential scatterer it has been shown that the characteristic scaling behavior of overlaps with system size is not only valid for the ground state wave function overlap, but also for low-lying excited states $|\varepsilon\rangle$~\cite{Ohtaka,Feldkamp}. Here, the energies $\varepsilon$ are measured relative to the ground state energy of the final Hamiltonian. For a finite-size system it has been shown by the authors in Ref.~\cite{Feldkamp} that within an interval $\Delta E=W/N$ (the single-particle level spacing) with $W$ the bandwidth the function $\sigma(E_n)$, $E_n=nW/N$, defined as $\sigma(E_n)=\sum_{\varepsilon=E_n}^{E_n+\Delta E}| \langle \varepsilon | 0 \rangle |^2$ has a scaling behavior similar to that in the Anderson orthogonality catastrophe, namely $\sigma(E_n)\sim (n/N)^{\alpha^2-1} \sim E_n^{\alpha^2-1}$ provided the energy $E_n$ is small. This scaling behavior in energy is intimately connected to the scaling behavior with system size in der Anderson orthogonality catastrophe, however, it is an extension to excited states. The asymptotic power-law behavior of $\mathcal{L}(t)$ is therefore not just a consequence of the vanishing ground state overlap, but rather due to the existence of a multitude of low-energy excitations satisfying the characteristic scaling behavior also found in the Anderson orthogonality catastrophe.

Due to the correspondence between absorption spectra and work distributions we know that the average energy in the system $E_f=\langle0 | H_f |0 \rangle$ after the switch on of the perturbation is larger than the ground state energy $E_f^{gs}$ of $H_f$. After the quench the system has (on average) an excess energy $w=E_f-E_f^{gs}$, in the context of the work distribution one can term $w$ the dissipated work. The asymptotic long-time behavior of the Loschmidt echo $\mathcal{L}(t)$ suggests that in course of time the system redistributes this excess energy completely into a multitude of low-energy excitations.

\section{Conclusion}

In this work we have discussed the X-ray edge singularity in optical spectra of quantum dots. We presented a general framework that allows to determine analytically the singular threshold behavior of absorption spectra in quantum dots at zero temperature. This establishes an important generalization of x-ray edge physics to experimentally accessible environments that can be used to observe x-ray edge physics in a controlled setup. Moreover, we highlighted the correspondence of the spectra to work distributions and to the Loschmidt echo. The presented framework might also be useful in other contexts such as decoherence in charge qubits.

\begin{acknowledgments} We acknowledge valuable discussions with J.~von Delft. We thank M.~Hanl and A.~Weichselbaum for providing us the NRG data. This work was supported by SFB~TR12 of the Deutsche Forschungsgemeinschaft (DFG), the Center for Nanoscience (CeNS) Munich, and the German Excellence Initiative via the Nanosystems Initiative Munich (NIM).
\end{acknowledgments}

\appendix

\section{Model and experiment}
\label{App_A}

In this appendix we present one possible absorption experiment whose effective description is governed by the Hamiltonians in Eq.~(\ref{eq_Hamiltonians}). Consider a single semiconductor quantum dot embedded in a Schottky diode structure. Such quantum dots coupled to a fermionic reservoir can generically be described by Anderson impurity models:
\begin{eqnarray}
	H & = & \sum_{k\sigma} \varepsilon_k c_{k\sigma}^\dag c_{k\sigma} + V \sum_{k\sigma} \left[ c_{k\sigma}^\dag d_\sigma + d_\sigma^\dag c_{k\sigma}\right] \nonumber \\
	  & & + \sum_\sigma \varepsilon_0 d_\sigma^\dag d_\sigma + U d_\uparrow^\dag d_\uparrow d_\downarrow^\dag d_\downarrow. 
\end{eqnarray}
The operator $c_{k\sigma}^\dag$ creates an electron in the reservoir with spin $\sigma$ and wave vector $k$ that is measured relative to $k_F$. The quantum numbers $k$ refer to an effective chiral one-dimensional description, see also the main text. The quantum dot is modeled by a single level with energy $\varepsilon_0$ in each spin channel. The operator $d_\sigma^\dag$ creates one electron of spin $\sigma$ on the dot. The two spin channels are coupled via the local onsite interaction of strength $U$. The hybridization $\Delta=\pi \rho_0 V^2$ with $\rho_0$ the noninteracting density of states constitutes a second important energy scale. For a semiconductor quantum dot in such a Schottky diode structure $\Delta$ can be tuned up to such large values that $U$ and $\Delta$ are of the same order. For $\Delta>U$ renormalization group studies reveal that the physical properties of the system are dominated by fixed points that correspond to the noninteracting limit of the above Hamiltonian with $U=0$~\cite{Krishnamurthy}. In this regime it is therefore valid to assume that the two spin channels are decoupled each of which can be modeled by a resonant level Hamiltonian $\tilde{H}$. Due to this decoupling we can restrict ourselves to a single channel of spinless electrons in the following:
\beq
	\tilde{H} = \sum_{k} \varepsilon_k c_{k}^\dag c_{k} +\varepsilon_0 d^\dag d + V \sum_{k} \left[ c_{k}^\dag d + d^\dag c_k \right].
\eeq
This is the effective description of the quantum dot before the absorption of a photon. As explained in the main text one effect of the absorption is the shift of the local level energy $\varepsilon_0 \to \varepsilon_0-U_{\textrm{\tiny eh}}$ via the attractive electron-hole interaction $U_{\textrm{\tiny eh}}$. Additionally, the absorption is associated with the switch on of a local potential scatterer for the electrons in the reservoir such that we have the following initial ($\mathcal{H}_i$) and final ($\mathcal{H}_f$) Hamiltonians:
\begin{eqnarray}
	\mathcal{H}_i & = & \sum_{k} \varepsilon_k c_{k}^\dag c_{k} +\varepsilon_0 d^\dag d + V \sum_{k} \left[ c_{k}^\dag d + d^\dag c_k \right] \nonumber \\
	\mathcal{H}_f & = & \sum_{k} \varepsilon_k c_{k}^\dag c_{k} +(\varepsilon_0-U_{\textrm{\tiny eh}}) d^\dag d \nonumber \\
			& & + V \sum_{k} \left[ c_{k}^\dag d + d^\dag c_k \right] -g\sum_{kk'} c_k^\dag c_{k'}. 
\end{eqnarray}
Let $W$ be the unitary transformation that diagonalizes $\mathcal{H}_i$, i.e.,
\beq
	H_i=W \mathcal{H}_i W^\dag=\sum_k \tilde{\varepsilon}_k c_k^\dag c_k + \varepsilon_i d^\dag d.
\eeq
where the matrix elements of $W$ are defined by the equations:
\begin{eqnarray}
	& & W d W^\dag=W_{dd} d + \sum_{k} W_{dk} c_k, \nonumber \\
	& & W c_k W^\dag=W_{kd} d + \sum_{k'} W_{kk'} c_{k'},
\end{eqnarray}
It is straightforward to show that all matrix elements have a square root scaling with system size, i.e. $W_{ll'}\sim L^{-1/2}$ with $l=k,d$. In the new basis the final Hamiltonian equals:
\begin{eqnarray}
	H_f & = & W \mathcal{H}_fW^\dag=\sum_k \tilde{\varepsilon}_k c_k^\dag c_k +  \varepsilon_f d^\dag d \nonumber \\ 
	    & &  +  \sum_k  \left[\tilde{V} c_k^\dag d+\tilde{V}^* d^\dag c_k \right] -  \sum_{kk'} \tilde{g}_{kk'} c_k^\dag c_{k'}.
\end{eqnarray}
where the new coupling constants are given in terms of the matrix elements of $W$ in the following way
\begin{eqnarray}
	& & \varepsilon_f=\varepsilon_i-g \left| \sum_{k} W_{k d} \right|^2 \nonumber \\
	& & \tilde{V}= -g \sum_k W_{kd} \nonumber \\
	& & \tilde{g}_{kk'}= -g \sum_{qq'} W^*_{qk} W_{q' k'} - U_{\textrm{\tiny eh}} W_{dk}^* W_{dk'}
\end{eqnarray}
Here, we have neglected all terms whose contribution vanishes in the thermodynamic limit. For the singular behavior of the absorption spectrum at the threshold only low-energy excitations are relevant. For the description of the low-energy sector one can replace the coupling constants  $\tilde{g}_{kk'} \to \tilde{g}_{00}$ by their values at the Fermi level. Concluding, we have shown one possible experimental scenario that leads to the model Hamiltonians in Eq.~(\ref{eq_Hamiltonians}).

\end{document}